\documentclass[11pt]{article}
\usepackage{url}
\begin{document}
\baselineskip=15pt
\bibliographystyle{plain}

\begin{center}
{\large {\bf Hamming's Original Paper Rewritten in Symbolic Form:\\
A Preamble to Coding Theory}}\\
 \vspace{1cm}
H. Gopalakrishna Gadiyar and R. Padma \\
School of Advanced Sciences, V. I. T. University, Vellore 632 014, India\\
E-mail: \{gadiyar, rpadma\}@vit.ac.in
\end{center}
\begin{abstract} In this note we try to bring out the ideas of Hamming's classic paper on coding theory in a form understandable by undergraduate students of mathematics.
\end{abstract}
\section{Introduction} Long ago Brinn \cite{brinn} made an appeal for introducing algebraic coding theory in the undergraduate curriculum. This goal is more urgent now than ever with the ubiquity of computers and communication devices. This article is still worth reading though it was written thirty years ago. 

The aim of this note is to write Hamming's paper in symbolic form. Hamming used the deceptively simple idea of interleaving parity check which helps to locate and hence correct errors. 

This note will enable readers to move on to the now classic books by V. Pless \cite{pless}and Berlekamp \cite{berlekamp}. Hamming's classic paper \cite{hamming} is difficult to read because the mathematics is written in words and tables. This was because the audience of mathematicians and engineers in those days who worked in applied fields preferred to avoid symbolic notation and algebra as far as possible. This situation has completely changed due to various reforms in the curriculum.

In the books by Birkhoff \cite{birkhoff} and Pless \cite{pless}, a more sophisticated approach is used where the group property of the codes is emphasized as this leads to the more recent developments. The parity check bits are not interleaving but an identity matrix appended to the end.

Our simple minded approach of translating Hamming's paper into symbolic form is to enable undergraduate students to understand the elegance and simplicity of Hamming's construction which is a mix of engineering thinking and mathematical thinking. The concepts of interleaving and parity check have their origin in engineering. The idea of coding belongs more to pure mathematics. Understanding Hamming's paper is essential for reading the book ``From error correcting codes through sphere packings to simple groups" by Thomas M. Thompson \cite{thompson} which is a delightful mix of history and pedagogy. This would enable undergraduate students to appreciate the unity of pure and applied mathematics, interdisciplinary and multidisciplinary research through this concrete example given in historical form. It would also enable them to take the more standard route for pursuing further developments in algebraic coding theory.  

\section{Hamming's construction}
Parity bits are an engineering trick to detect errors in a string of $0$'s and $1$'s. In its simplest form the number of $1$'s is counted and then computed modulo 2. The answer would be $0$ if the number of $1$'s is even and $1$ if the number of $1$'s is odd. This is appended to one end of the binary string. In Hamming's case he interleaves the parity check bits in a clever way for error correction. This is a conceptual leap beyond error detection which was well known then. 

Let $m$ be the number of information bits, $k$ the number of error correction bits and $n=m+k$. Since any $k$ bits represent numbers from $0$ to $2^k-1$, we need the condition that $2^k-1\ge n=m+k$. Hence if a single error has occurred, one can determine its position from the $k$-bit binary representation of its position number (Hamming calls it Checking number.) Hamming interleaves the $k$ check bits in positions $x_{2^0}, x_{2^1},\cdots x_{2^{k-1}}$. He places the $m=n-k$ information bits at the remaining positions. Hamming analyzed the case of $(7,4)$ code with the rate $\frac{4}{7}\sim 0.571$ in the modern notation. In the notation given above, $k=3$, $m=4$ and $n=7$.

The $k$ check bits are calculated as follows. At the encoding end, $x_1=x_{2^0}$ is determined by the partial parity check equation
\begin{equation}
x_1+x_3+x_5+x_7+\cdots =0
\end{equation} 
Notice that all these bits have their position numbers $1, 3, 5, 7 \cdots $ which when they are written in their binary representation have the least significant bit equal to 1. 
Hence if the single error has occurred in any one of the odd positions, then at the decoding end, the partial parity check equation will give
\begin{equation}
x_1+x_3+x_5+x_7+\cdots =1
\end{equation}  
Next, $x_2$ is determined (at the encoding end) by the equation 
\begin{equation}
x_2+x_3+x_6+x_7+\cdots =0
\end{equation} 
Notice that the binary representations ($10, 11, 110, 111, \cdots $) of the position numbers of $2, 3, 6, 7, \cdots $ have $1$ as their second bit from the right and $2$ is the smallest of these numbers. Hence if the single error has occurred in a position whose second bit is $1$, then at the decoding end, we would get
\begin{equation}
x_2+x_3+x_6+x_7+\cdots =1
\end{equation} 
Similarly $x_{2^2},\dots x_{2^{k-1}} $ are determined by the corresponding partial parity check equations.
Notice that $1, 2, 4, 8, \cdots , 2^{k-1}$ ($1, 10, 100, 1000, \cdots 100\cdots 0$)are the smallest numbers having $1$ in the first, second, third, fourth, $\cdots k^{th}$ positions in their binary representations. Thus the position number of the error bit is determined bit by bit from right to left. The least significant bit is zero if (1)is true and $1$ if (2) is true. Similarly the previous bit is zero if (3) is true and $1$ if (4) is true and so on. Once the position of the error bit is found, the bit can be corrected as a bit can take only two values: $0$ or $1$. 

At this point we would encourage the reader to look at the classic paper of Hamming paper \cite{hamming} which is freely down loadable from the Internet and then read the standard books listed below. 

\section{Pedagogical and historical comments}  \cite{tremblay} uses symbolic notation for bits with parity check matrix but the idea of interleaving is missed. \cite{huffman} also does not talk about interleaving parity check bits. \cite{birkhoff} and \cite{tremblay} discuss Hamming codes as a special case of group codes.


\begin{thebibliography}{10}
\bibitem{berlekamp} E. R. Berlekamp, {\it Algebraic coding theory}, McGraw - Hill, 1968.
\bibitem{berlakamp2} E. R. Berlekamp, {\it Key papers in the the development of Coding theory},  Ed: E. R. Berlekamp, IEEE Press, 1974.
\bibitem{birkhoff} G. Birkhoff and T. C. Bartee, {\it Modern applied algebra}, McGraw - Hill, 1970.
\bibitem{blake} Ian F. Blake, {\it Algebraic Coding theory: History and Development}, Dowden, Hutchinson \& Ross, 1973 
\bibitem{brinn} L. W. Brinn, {\it Algebraic coding theory in the undergraduate curriculum}, American Math. Monthly, {\bf 91, 8} October, 1984,  509-513.
\bibitem{hamming} R. W. Hamming, {\it Error detecting and error correcting codes}, The Bell System Technical Journal, {\bf 29} April 1950, 147-160. \url{http://www3.alcatel-lucent.com/bstj/vol29-1950/articles/bstj29-2-147.pdf}
\bibitem{hill} R. Hill, {\it A first course in coding theory}, Oxford University Press, 1986.
\bibitem{huffman} W. C. Huffman and V. Pless, {\it Fundamentals of error correcting codes}, Cambridge University Press, 2003. 
\bibitem{pless} V. Pless, {\it Introduction to the theory of error correcting codes}, Wiley - Interscience Series in Discrete Mathematics and Optimization, 1998.
\bibitem{thompson} T. M. Thompson, {\it From error correcting codes through sphere packings to simple groups}, Cambridge University Press, 1983.
\bibitem{tremblay} J. P. Tremblay and R. Manohar, {\it Discrete Mathematical Structures with applications to computer science}, McGraw-Hill Interamericana, 1975.

\end{thebibliography}
\end{document}